\documentclass[aps,prl,twocolumn,groupedaddress]{revtex4}
\usepackage{graphicx}

\begin{document}


\title{Observation of a subgap density of states in superconductor-normal metal bilayers\\
 in the Cooper limit}

\author{Zhenyi Long, M. D. Stewart, Jr.}

\author{Taejoon Kouh} 
 \altaffiliation[Presently at: ]{Aerospace and Mechanical Engineering Dept, Boston University, Boston, MA 02215.}

\author{James M. Valles, Jr.}
 \email{valles@physics.brown.edu}

\affiliation{Department of Physics, Brown University, Providence, RI 02912 }

\date{\today}

\begin{abstract}
We present transport and tunneling measurements of Pb-Ag bilayers with thicknesses, $d_{Pb}$ and $d_{Ag}$, that are much less than the superconducting coherence length. The transition temperature, $T_c$, and energy gap, $\Delta$, in the tunneling Density of States (DOS) decrease exponentially with $d_{Ag}$ at fixed $d_{Pb}$.  Simultaneously, a DOS that increases linearly from the Fermi energy grows and fills nearly 40\% of the gap for $T_c\approx$0.1 $T_{c,bulk}^{Pb}$. This behavior suggests that a growing fraction of quasiparticles decouple from the superconductor as $T_c\rightarrow $0.  The linear dependence is consistent with the quasiparticles becoming trapped on integrable trajectories in the metal layer.
  
\end{abstract}

\pacs{}

\maketitle

Simple metallic phases have a finite resistance in the zero temperature limit and a nonzero Density Of electronic States (DOS) at the Fermi energy, $E_F$. In two dimensions (2D), however, the scaling theory of localization asserts that simple metallic phases do not exist \cite{Abrahams}.  Ultrathin films exhibit a continuously decreasing conductance with decreasing temperature in support of this assertion \cite{Bergmann}. Nevertheless, there exist an increasing number of quasi-2D systems with metallic transport properties at low temperatures. These include 2D electron gases formed in semiconductor heterostructures, which appear to show an Insulator to Metal transition with electron density \cite{Kravchenko} and ultrathin films of metals balanced on the brink of a superconducting transition by disorder \cite{Chervenak1999} or magnetic field \cite{Goldman,Mason}. These metallic phases are probably not simple Fermi liquids and their existence depends on electron-electron interaction effects\cite{Kravchenko,Phillips}. Recently, two groups proposed that 2D arrays of superconducting islands immersed in a metal undergo a quantum superconductor to metal transition (SMT) with decreasing island concentration \cite{Feigelman,Spivak}. The resulting metallic phase has non-Fermi liquid properties including a pseudogap and anomalous magnetoresistance \cite{Spivak}.

Furthermore, theories of mesoscopic Superconductor-Normal metal (SN) structures have revealed mechanisms by which a finite DOS can appear within the energy gap of superconducting structures. These states, which give the DOS a hybrid superconductor-metal appearance, correspond to quasi-particles that become partially trapped in the N regions \cite{Ostrovsky2002,Melsen,Lodder}. Mesoscopic spatial fluctuations in the local conductivity can give rise to ``quasi-localized" states within the N region. These states appear within the gap, smearing the gap edge and creating a DOS down to $E_F$ \cite{Ostrovsky2002}. Within semi-classical models \cite{Melsen,Lodder,Schomerus}, the DOS depends on whether the dynamics in the N region is chaotic or integrable. Quasiparticles on ballistic, integrable trajectories can become quasi-trapped in N regions and contribute a subgap DOS that grows linearly from $E_F$. 

We have conducted a series of experiments on ultrathin SN bilayers in an effort to observe the proposed SMT \cite{Feigelman,Spivak} and its associated metallic phase.  These bilayers can be driven toward the metallic phase by increasing the metal layer thickness, $d_N$, at fixed superconductor thickness, $d_S$.  Within quasiclassical proximity effect theories, the bilayers ought to have a hard, BCS gap in the DOS and make a transition to a metal phase only as $d_N\rightarrow\infty$. Contrary to this expectation, our previous work showed that their superconducting transition temperatures, $T_c$, decrease faster with $d_N$ than quasi-classical predictions suggesting the approach to a SMT \cite{Kouh}. 
 
\begin{figure}
\includegraphics{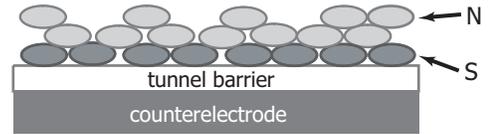}
\caption{Schematic cross section of the bilayer tunnel junction.}
\end{figure}

Here we present electron tunneling measurements showing that the DOS of ultrathin SN bilayers develops a hybrid superconductor-metal appearance that becomes more metallic as $T_c$ decreases.  Specifically, the superconducting gap of ultrathin Pb-Ag bilayers systematically fills with states as $T_c$ is decreased by increasing the metal thickness, $d_{Ag}$ at fixed superconductor thickness, $d_{Pb}$. The subgap DOS is finite at $E_F$ and rises linearly with energy with a slope that increases with $d_{Ag}$. Recent theories suggest that these subgap states are quasiparticles that become decoupled from the superconductor for times longer than the superconducting coherence time \cite{Ostrovsky2002,Melsen,Lodder,Schomerus}. The linear dependence is consistent with semiclassical theories of quasiparticles becoming trapped on integrable trajectories\cite{Melsen,Lodder,Schomerus}.   

For these studies, Pb/Ag bilayers with ultraclean Pb/Ag interfaces were fabricated and measured {\it in situ} using quench condensation techniques in the UHV environment of a dilution refrigerator cryostat\cite{Kouh}.  The metals were thermally evaporated onto fire polished glass substrates held at 8 K.  Au/Ge contact pads and oxidized Al counterelectrodes with a small amount of magnetic impurities to prevent them from superconducting were deposited prior to cryostat mounting.  To form bilayers, a thin $<$ 6 nm, electrically discontinuous, Pb film was deposited first followed by, without breaking vacuum or warming, a series of Ag depositions (see Fig. 1).  The latter drove the bilayer through an insulator to superconductor transition \cite{HsuProximity,Merchant}.  This procedure yielded a series of bilayers with a single $d_{Pb}$ and a range of $d_{Ag}$ that were probed with the same, 1.25 mm$^2$ area, tunneling counterelectrode and barrier.  The transport and tunneling measurements were performed using standard 4 terminal, low frequency AC techniques. Data acquired on a series of bilayers with $d_{Pb} =$ 4.0 nm and 4.2 $ < d_{Ag} < $ 19.3 nm are presented here.  This data set is the most complete and systematic. It exhibits features that are similar to those of other series with  1.5 $< d_{Pb} <$ 6.0 nm. 
 
\begin{figure}
 \includegraphics{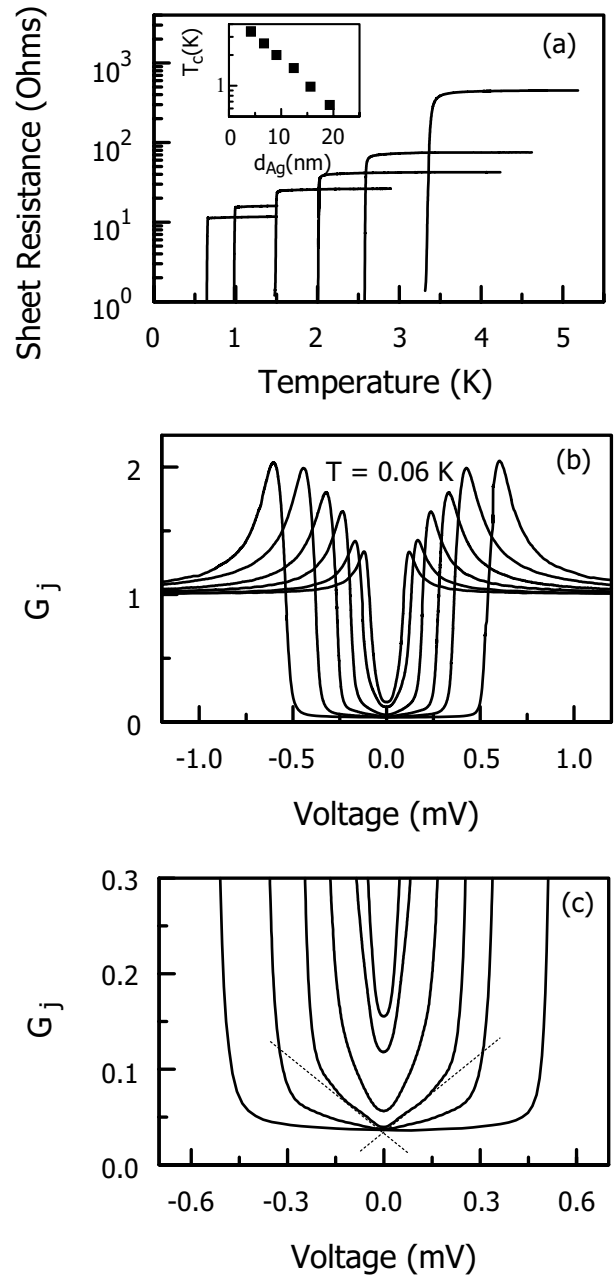}
 \caption{(a) Sheet resistance vs. temperature for Pb/Ag bilayers with $d_{Pb}=$ 4 nm and  $d_{Ag}=$ 4.2, 6.7, 9.1, 12.4, 15.6, 19.3 nm. Inset: Semilog plot $T_c$ vs. $d_{Ag}$. (b)  $G_j$  vs. $V$ for the same bilayers. (c) Same data as in (b) on a finer voltage scale. The dashed lines are
guides for the eye.}
\end{figure}

\begin{figure}
 \includegraphics{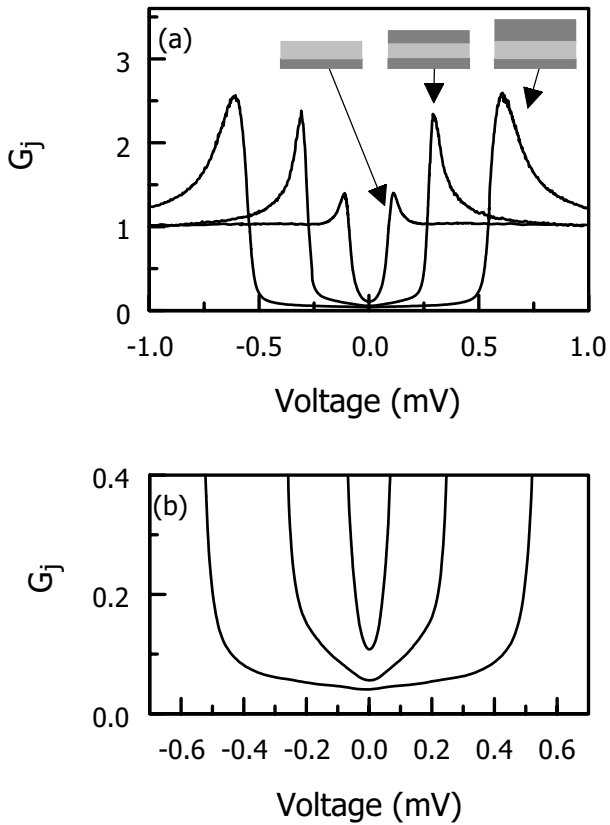}
 \caption{(a)  $G_j$ at $T =$ 60 mK vs. voltage for a bilayer $d_{Pb}/d_{Ag} =$ 1.4 nm/7.1 nm, and two trilayers, $d_{Pb}/d_{Ag}/d_{Pb}$, of 1.4 nm/7.1 nm/2.2 nm and 1.4 nm/7.1 nm/5.4 nm.  (b) Same data as in (a) on finer scales.}
 \end{figure}

The normalized, superconducting bilayer DOS, $N_S(E)$, is obtained from the normalized differential conductance, $G_j$, of the tunnel junction\cite{Tinkham}: 
$$G_{j}=\frac{G_S}{G_N}=-\int^\infty_{-\infty}{N_{S}(E)}{\frac{\partial{f(E+eV)}}{\partial{(eV)}}}dE $$
where $G_S$ and $G_N$ are the differential conductances, $dI/dV$, in the superconducting and normal states, respectively. $I$ is the tunnel current and $V$ is the voltage across the junction.  $f$ is the Fermi function and $E$ is the energy measured from $E_F$.  At low  temperatures, $T <$ 0.1 $T_c$, $G_{j}\approx{N_S(eV)}$. In the Cooper limit \cite{Cooper,DeGennes,Fominov}, $N_S$ is predicted to assume the BCS form, $N_S^{BCS}(E,\Delta)\!=\!Re(E/\sqrt{(E^2-\Delta^2)})$ where $\Delta$ is the energy gap.  

The evolution of the resistive transitions, $R(T)$ and tunneling conductances, $G_j$, of a bilayer series driven toward the metallic state is shown in Fig. 2.  The transitions are sharp and $T_c$, defined as the temperature at which $R(T)$ is half its normal state value, drops exponentially with $d_{Ag}$ (inset). The $G_{j}$, obtained at $T =$ 60 mK $\ll T_c$, qualitatively resemble the BCS form, exhibiting an energy gap structure consisting of symmetric peaks and a depression at low voltages (Fig. 2b).  Within the gap region, however, $G_j$ has an approximately linear voltage dependence and a finite value at zero voltage (Fig. 2c) rather than an exponentially small value.  This subgap conductance grows with $d_{Ag}$ and fills nearly 40\% of the gap for the bilayer with $T_c =$ 0.67 K.  In addition, the conductance peaks are shorter and broader than the BCS prediction.

Depositing Pb atop a bilayer to create a trilayer reverses the above evolution as shown in Fig. 3 for a bilayer with $d_{Pb} =$ 1.4 nm and $d_{Ag} =$ 7.1 nm.  It's $G_j$ resembled that of the lower $T_c$ bilayers in Fig. 2.  Adding an upper Pb layer sharpened the peaks, increased the energy gap and reduced the slope and zero voltage bias value of $G_j$.  A second Pb evaporation continued these trends.   We hasten to note that the reduction in the subgap conductance induced by the upper Pb layer is a sign that the subgap conductance reflects an intrinsic feature of the DOS.  A subgap conductance stemming from leakage would be unaffected by an upper Pb layer. 

Except for the subgap conductance and the broadened conductance peaks, the data in Fig. 2 follow quasiclassical models of the proximity effect in the Cooper limit  \cite{Cooper, DeGennes, Fominov, Bourgeois}. In this limit, which applies to bilayers with $d_S, d_N<< \xi$, the superconducting coherence length, electrons near $E_F$ pass back and forth between the S and N regions rapidly compared to $\Delta/\hbar$.  The effective superconducting coupling constant, $\lambda\propto{ln(T_c)}\propto{ln(\Delta)}$, is the volume average of the coupling constants in the N and S regions and thus, depends linearly on $d_{Ag}$ in agreement with the inset of Fig. 2a \cite{Cooper}.   The rapid motion renders the pairing amplitude uniform across the bilayer and thus, leads to a BCS form for the tunneling DOS \cite{DeGennes} that roughly agrees with the data. 

The extra breadth in the peaks can be attributed to the existence of a distribution of energy gaps in the bilayers.  A distribution can naturally arise due to spatial variations in the film thickness ratio $x=d_{Ag}/d_{Pb}$ since $\Delta\propto T_c$ depends exponentially on $x$.  Presuming $\Delta=\Delta_{0}exp(-kx)$, a (random) normal distribution of $x$ yields a log-normal distribution for $\Delta$ and a broadened form of $N_S(E)$:
$$N_{S}^\sigma(E,\overline{\Delta})=\!\!\\\frac{1}{\sqrt{2\pi}k\sigma}\!\!\int_0^{\Delta_0}\!\!\!N_{S}^{BCS}(E,\Delta)exp(-\frac{(ln(\frac{\overline{\Delta}}{\Delta}))^2}{2(k\sigma)^2})\frac{d\Delta}{\Delta}$$
where $\overline{\Delta}$ is the most probable energy gap and $\sigma$ is the width of the $x$ distribution.  The parameters $\Delta_0 = $0.88 meV and $k =$ 0.46 were obtained by estimating $\Delta$ for each bilayer as the voltage at which $G_j=$1 and fitting to the exponential form. Calculated $G_j$ (see Fig. 4a) with broadened peaks that resemble the data require $\sigma < $25\% of $x$. 

The states that systematically fill the superconducting gap, however, do not readily conform to a gap distribution model.  The log-normal gap distribution drops too rapidly below $\overline{\Delta}$  to reproduce the linear DOS and finite zero bias conductance(see Fig. 4a). Creating the observed DOS at $E_F$ would require that $\sigma$ assume values at least 10 times larger than the values used to fit the peaks. Thus, we are led to the conclusion that the subgap DOS corresponds to quasiparticles that fall outside the Cooper limit picture and thus must relate to a new physical mechanism. A similar problem was encountered by Gupta and coworkers\cite{Gupta} who were unable to fit subgap structure in their STM data on Au/Nb films.
 
\begin{figure}
 \includegraphics{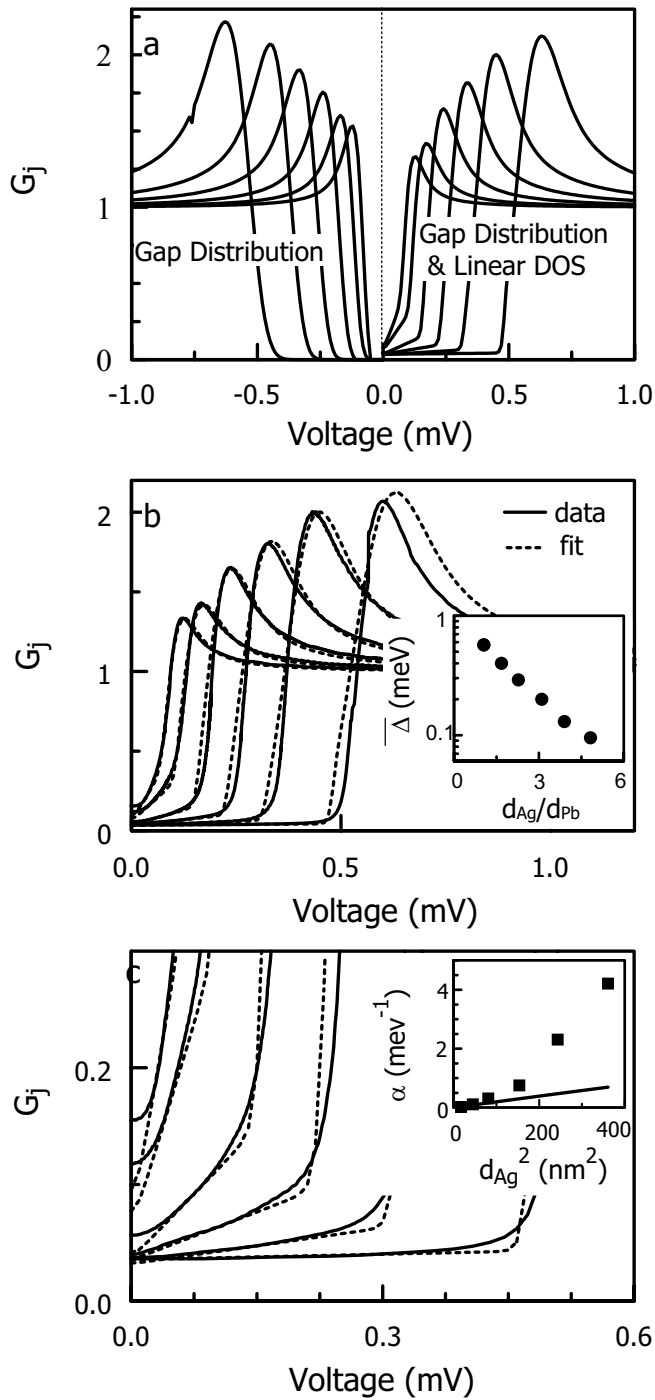}
 \caption{(a) Comparison of $G_j$ calculated using a log-normal gap distribution (left) and a log-normal distribution combined with a linear DOS within the gap (right). The corresponding curves have the same fitting parameters, $\overline{\Delta}$ and $\sigma$. (b) Data from Fig. 2a with fits. The fitting parameter $\sigma/(\frac{d_{Ag}}{d_{Pb}})$ decreases from 0.23 to 0.11 as $d_{Ag}$ goes from 4.2 nm to 19.3 nm, while $\beta$ increases from 0.033 to 0.065. Inset: Semilog plot of $\overline{\Delta}$, vs. $d_{Ag}/d_{Pb}$. (c) Same as (b) on finer scales.  Inset: $\alpha$ vs. $d_{Ag}^2$.  The solid line gives the prediction based on transport measurements.}
 \end{figure}

According to semi-classical theories, subgap states correspond to quasiparticles that propagate in an N region over distances more than $\xi$ between successive Andreev scattering events with an NS interface\cite{Usadel,Belzig,Melsen,Lodder}. This situation normally occurs in SN structures with N region dimensions exceeding $\xi$, \cite{Belzig,Moussy,Gueron,Truscott,Tessmer} or structures with integrable dynamics \cite{Melsen,Lodder}, in which quasiparticles can execute nearly closed trajectories.  Interestingly, the predicted DOS grows approximately linearly with energy for a variety of structures\cite{Belzig,Melsen,Lodder,Schomerus}. In particular, Melsen and coworkers \cite{Melsen} proposed that for a rectangular integrable N billiard attached to a bulk S region, the distribution of path lengths, $P(L)$, between successive Andreev scattering events follows $P(L)\propto 1/L^{3}$ as $L \rightarrow \infty$.  The normalized slope of the resulting linear DOS is $2/(\pi E_{Th}) $ \cite{Melsen}, where $E_{Th}$ is the Thouless energy. Furthermore, the more general case of a mixed phase space that consists of both regular (integrable) and chaotic regions also yields a linear DOS and, in addition, a constant background DOS arising from phase space regions that are completely disconnected from the superconductor \cite{Schomerus}.

Guided by the above considerations and low temperature STM measurements showing that the bilayers consist of crystalline Pb and Ag grains\cite{Ekinci,Long}, we now assume that the bilayers present a mixed phase space of regular and chaotic regions\footnote{The quasiparticle propagation in these granular bilayers is neither strictly ballistic nor strictly diffusive. It is generally believed that intragrain propagation is ballistic while the intergrain propagation is diffusive.}. Quasiparticles with chaotic trajectories contribute a BCS like portion, $N_{S}^\sigma$, to the DOS while those with regular trajectories contribute a linear subgap DOS. Thus, we parameterize $N_S(E)$:
 $$N_S(E)=\cases{N_{S}^\sigma(E,\overline{\Delta}) &if $E>E_{c}$\cr
\alpha{E}+\beta &if $E\leq{E_{c}}$}$$ 
where $\alpha$ and $\beta$ are the slope and the intercept of the linear dependence, respectively, and $E_{c}$ is the energy at which the two DOS intersect. Fig. 4a(right) shows $G_j$ calculated from this DOS \footnote{To ensure conservation of states for this form, we reduced $N_{S}^\sigma$ by the fraction of subgap states. This reduction made the peak heights on the left and right hand sides of Fig. 4a differ.}. Fits to the data optimized to capture the subgap slopes and the peak heights are shown in Fig. 4b and Fig. 4c. ${\overline{\Delta}}$ decreases exponentially with $d_{Ag}$ (inset Fig. 4b) consistent with expectations for the Cooper limit. Since $\alpha$ is the slope of the DOS averaged over the regular phase space regions, $\alpha$ is proportional to the average $\overline{E_{Th}^{-1}}$. For these bilayers, we expect $\overline{E_{Th}^{-1}}\sim d_{Ag}^2/(\hbar D)$, where $D$ is the intergrain diffusivity and thus $\alpha\propto d_{Ag}^2$.  Fig. 4c compares the experimental $\alpha$ values with those estimated by $2d_{Ag}^2/(\pi\hbar D)$, using $D=5\times 10^{-3}m^2s^{-1}$ as obtained from transport measurements.  The two roughly agree at small $d_{Ag}$ and deviate at higher $d_{Ag}$ where the experimental $\alpha$ values grow more rapidly than expected. The deviation suggests that $\alpha$ also depends on $T_c$.  This case has not been considered by the current theories \cite{Melsen, Schomerus}. The trilayer data support this conjecture, exhibiting a reduction in $\alpha$ when $T_c$ is increased at fixed $d_{Ag}$.  Regardless, this deviation implies that the regular phase space regions occupy an increasing fraction of the total phase space volume as $d_{Ag}$ increases. The concomitant increase of $\beta$ implies a growing fraction of disconnected regions, as well. Consequently, as $T_c$ decreases, a growing fraction of quasiparticles decouple from the S layer for times much greater than the superconducting coherence time.   

Qualitatively, the growth in the subgap DOS gives the DOS a hybrid-metal-superconductor appearance that may be a sign of an approaching SMT.  Ultrathin films near the disorder tuned superconductor-insulator transition exhibit a similar filling of the gap \cite{Valles1992,Hsu1995} and low energy states have been invoked as a source of dissipation that drives the Quantum SMT observed in nanowires \cite{Bezryadin,Sachdev,Tewari}.  Perhaps these states arise due to similar quasiparticle trapping effects. 
 
Tunneling experiments on ultrathin Pb-Ag bilayers at low reduced temperatures ($T/T_c <$ 0.1) have revealed an unexpected linear DOS within their superconducting energy gap.  The fraction of the gap filled with states grows with increasing $d_{Ag}$.  We have identified the subgap states as quasiparticles that are weakly coupled to the superconductor layer.  Within a semiclassical picture, these quasiparticles become trapped in regular regions of phase space. 

\begin{acknowledgments}
The work was supported by NSF-DMR0203608. We acknowledge helpful conversations with Dmitri Feldman.
\end{acknowledgments}

\end{document}